\documentclass[reqno,12pt]{amsart}
\usepackage{url}

\usepackage{graphicx}
\usepackage{epstopdf}
\usepackage{ljm-auth}
\usepackage{cite}
\usepackage{enumitem}
\usepackage{hyperref}

\author{Leonid B.\,Sokolinsky}
\crauthor{L.B. Sokolinsky} 

\tit{Analytical Estimation of the Scalability of Iterative Numerical Algorithms on Distributed Memory Multiprocessors}
\shorttit{Analytical estimation of scalability} 

\begin{document}
\maketit

\address{South Ural State University, Chelyabinsk, Russia}

\email{Leonid.Sokolinsky@susu.ru}

\Received{November 15, 2017}

\abstract{This article presents a new high-level parallel computational model named BSF -- Bulk Synchronous Farm. The BSF model extends the BSP model to deal with the compute-intensive iterative numerical methods executed on distributed-memory multiprocessor systems. The BSF model is based on the master-worker paradigm and the SPMD programming model. The BSF model makes it possible to predict the upper scalability bound of a BSF-program with great accuracy. The BSF model also provides equations for estimating the speedup and parallel efficiency of a BSF-program.}

\raggedbottom

\notes{0}{\emergencystretch=20pt
\thank{This research was partially supported by the Russian Foundation for Basic Research (project No.~17-07-00352a), by the Ministry of Education and Science of Russian Federation (gov.~order No.~2.7905.2017/8.9) and by the Government of the Russian Federation according to Act 211 (contract No.~02.A03.21.0011)}}

\section{Introduction}
One of the most important properties of a numerical algorithm designed for large-scale cluster systems is scalability. \emph{Scalability} can be defined as a measure of a parallel system's capacity to decrease computation time in proportion to the number of processors. The upper bound of scalability is an integral characteristic of a parallel algorithm/program. The \emph{upper bound of scalability} is the least number of processor nodes for which the speedup takes the maximal value. It is valuable to be able to estimate the upper bound of scalability in early phases of program development; the parallel computation model is a tool providing this possibility. A \emph{model of computation} is a framework for specifying and analyzing algorithms or programs~\cite{Bilardi2011}. Many parallel computation models have been proposed for distributed-memory multiprocessors. The most famous of these models are the \emph{BSP model family} (see~\cite{Valiant1990, Auf-der-Heide2001, Valiant2011, Blanco2004, Gerbessiotis2015, Cha2001}) and the \emph{LogP model family} (see~\cite{Culler1993, Alexandrov1997, Liu2012, Lu2015, Ino2001, Cameron2007, Yuan2010}). Most of these models are low-level models and require detailed description of the structure of the algorithm to the level of code in a programming language or pseudocode. This article extends the basic BSP (Bulk Synchronous Parallelism) model~\cite{Tiskin2011} to deal with the compute-intensive iterative numerical methods executed on distributed-memory multiprocessor systems. Iterative methods are an important class of numerical methods. An overview of various iterative methods can be found in~\cite{Hageman1981, Kelley1995, Hadjidimos1987}. The new parallel computation model proposed in this article was named \emph{BSF -- Bulk Synchronous Farm}. The BSF model is a high-level parallel computation model based on the master-worker (master-slave) framework~\cite{Sahni1996} and the SPMD (Single-Program-Multiple-Data) programming model~\cite{Darema1988, Darema2011}. A distinctive feature of the BSF model is the ability to estimate the upper bound of scalability in the early stages of the algorithm design.

The rest of the article is organized as follows. In Section \ref{BSF-model}, the BSF parallel computation model presented in this paper is described.  Section \ref{BSF-scalability} introduces a cost metric for BSF-programs and provides equations for estimating the speedup and parallel efficiency of an algorithm before its implementation in a programming language. Moreover, a simple inequality to estimate the upper scalability bound of a BSF-program is deduced. Section \ref{Conclusion} summarises the results and outlines some directions for future research.

\section{BSF computational model}\label{BSF-model}

\begin{figure}
\includegraphics[scale=1.0]{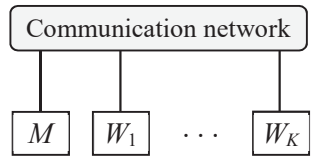}
\caption{BSF-computer structure. $M$ -- master node; $W_1, \ldots, W_K$ -- worker nodes.}\label{Fig-BSF-computer}
\end{figure}

The \emph{BSF} (\emph{Bulk Synchronous Farm}) model is intended for multiprocessor systems with distributed memory. A \emph{BSF-computer} consists of a collection of homogeneous computing nodes with private memory connected by a communication network delivering messages among the nodes. There is just one node called the \emph{master-node} in a BSF-computer. The rest of the nodes are the \emph{worker-nodes}. A BSF-computer must include at least one master-node and one worker-node. The BSF-computer layout is shown in Fig.~\ref{Fig-BSF-computer}.

\begin{figure}
\includegraphics[scale=1.0]{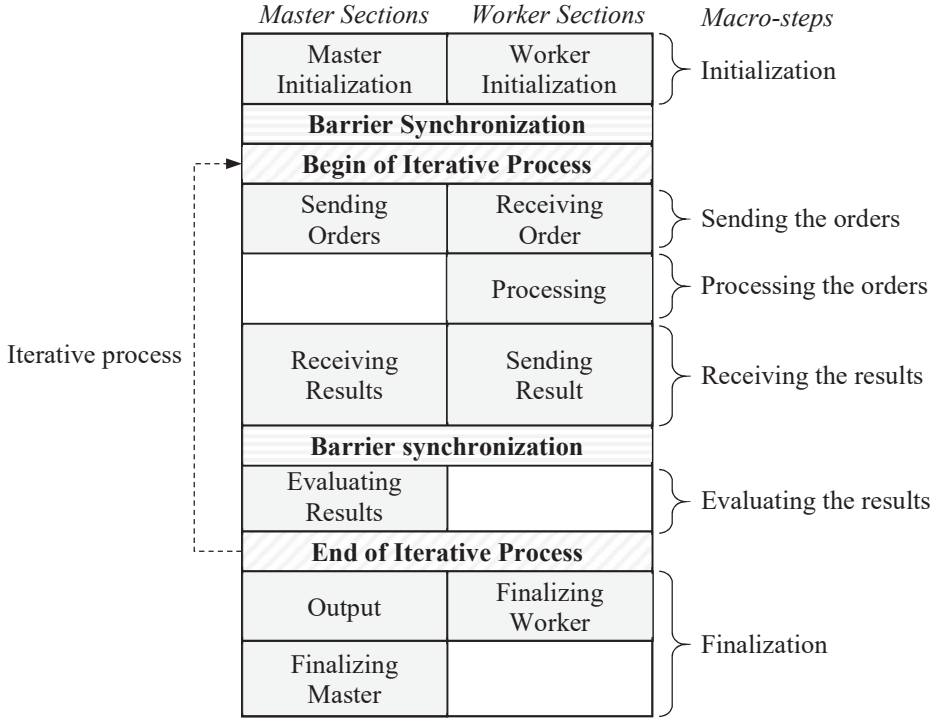}
\caption{BSF-program structure.}\label{Fig-BSF-program}
\end{figure}

A BSF-computer utilizes the \emph{SPMD} programming model according to which all the worker-nodes executes the same program but process different data. A BSF-program consists of sequences of macro-steps and global barrier synchronizations performed by the master and all the workers. Each macro-step is divided into two sections: the master section and the worker section. The master section includes instructions performed by only the master. A worker section includes instructions performed by only the workers. The sequential order of the master section and the worker section within the macro-step is not important. All the worker nodes operate on the same data array, but the base address of the data assigned to the worker-node for processing is determined by the logical number of this node. A BSF-program includes the following sequential sections (see Fig.~\ref{Fig-BSF-program}):
\begin{itemize}
  \item initialization;
  \item iterative process;
  \item finalization.
\end{itemize}

\emph{Initialization} is a macro-step in which the master and workers read or generate input data. Initialization is followed by barrier synchronization. The \emph{iterative process} repeatedly performs its body until the exit condition checked by the master becomes true. In the \emph{finalization macro-step}, the master outputs the results and ends the program.

The \emph{body of the iterative process} includes the following macro-steps:
\begin{enumerate}[label=\arabic*)]
  \item sending orders (from master to workers);
  \item processing orders (by workers);
  \item receiving results (from workers to master);
  \item evaluating the results (by master).
\end{enumerate}

\begin{figure}
\includegraphics[scale=0.7]{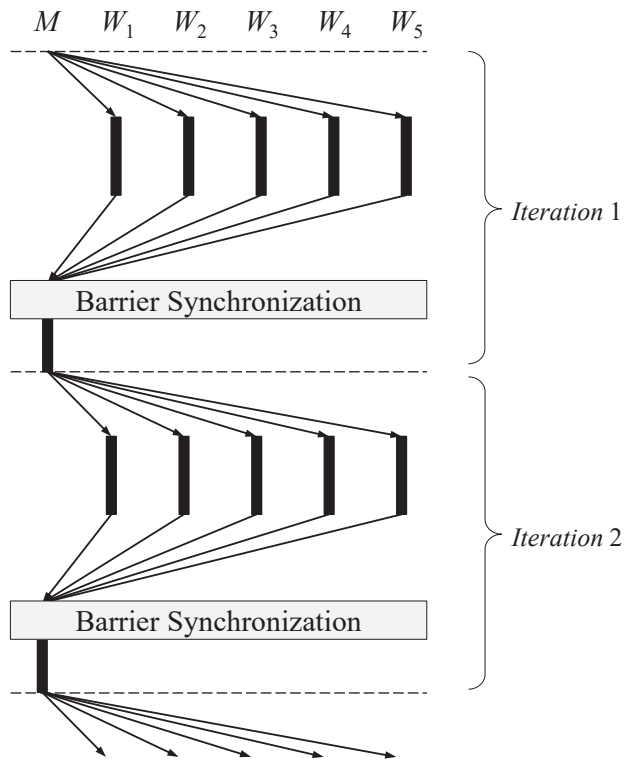}
\caption{BSF computation framework with one master $M$ and five workers $W_1, \ldots, W_5$ (the thick vertical lines denote local computations, the thin lines with arrows -- data transfers, and the horizontal dashed lines -- borders between iterations).}\label{Fig-BSF-computation}
\end{figure}

In the first macro-step, the master sends the same orders to all workers. Then, the workers execute the received orders (the master is idle at that time). All the workers execute the same program code but operate on different data with a base address which depends on the worker-node number. Therefore, all workers spend the same amount of time on calculation. There are no data transfers between nodes during order processing. In the third step, all workers send the results to the master. Next, global barrier synchronization is performed. During the fourth step, the master evaluates the results it has received. The workers are idle at this time. After evaluation of the results, the master checks the exit condition. If the exit condition is true, then the iterative process is finished, otherwise the iterative process is continued. BSF-program execution is illustrated in Fig. \ref{Fig-BSF-computation}.

\section{Evaluation of BSF-program scalability}\label{BSF-scalability}
The main characteristic of scalability is the speedup. For a parallel program, \emph{speedup} $a(K)$ can be defined as a ratio of execution time $T_1$ on one computing node to execution time $T_K$ on $K$ computing nodes:
\begin{equation} \label{Eq-speedup}
a(K)=\frac{T_1}{T_K}.
\end{equation}
Parallel efficiency is another important characteristic of scalability. \emph{Parallel efficiency} $e(K)$ can be defined as a ratio of speedup $a(K)$ to the number $K$ of processors:
\begin{equation} \label{Eq-efficiency}
e(K)=\frac{a(K)}{K}.
\end{equation}

This section offers a cost metric which can be used to estimate the scalability of a BSF-program. We assume that time spent on initialization and finalization of a BSF-program is negligible compared to the cost of iterative process execution. The cost of an iterative process is equal to the sum of the costs of separate iterations. Therefore, to estimate the execution time of a BSF program, it is sufficient to obtain an estimation of the execution time of a single iteration. For this purpose, the following main parameters of the BSF model are introduced:
\begin{itemize}
  \item [$K$:] the number of worker-nodes;
  \item [$L$:] an upper bound on the latency, or delay, incurred in communicating a message containing one byte from its source node to its target node;
  \item [${{t}_{s}}$:] the time that the master-node is engaged in sending one order to one worker-node, excluding latency;
  \item [${{t}_{w}}$:] the time a \mbox{BSF-computer} with one worker-node needs to perform one order;
  \item [${{t}_{r}}$:] the total time that the master-node is engaged in receiving the results from all worker-nodes, excluding latency;
  \item [${{t}_{p}}$:] the total time that the master-node is engaged in evaluating the results received from all worker-nodes.
\end{itemize}
The global barrier synchronization performed in iterative process is implemented by the master waiting for completion of reading all messages from workers, and therefore, it does not require an additional cost.

The time $T_1$ needed for the execution of a single iteration by a BSF-computer with one master-node and one worker-node can be calculated as follows:
\begin{equation}\label{Eq-T_1_a}
T_1 = t_s + t_w + L + t_p + t_r + L,
\end{equation}
which is equivalent to
\begin{equation}\label{Eq-T_1}
T_1 = 2L+t_s + t_w + t_p + t_r.
\end{equation}

Now, let us calculate the time $T_K$ a \mbox{BSF-computer} with one master-node and $K$ worker-nodes needs to execute a single iteration. All of the workers receive the same message from the master, so the total time for sending messages from the master to the workers is equal to $K(L + t_s)$. All of the workers perform the same program code on their own data segment, so the time of order execution by a group with $K$ workers is equal to $t_w/K$. The resulting data volume produced by the workers is a parameter of the task and does not depend on $K$, so the total time needed for sending messages from the workers to the master is equal to $K \cdot L + t_r$. The time needed for the master to process the results received from the workers is also a task parameter and does not depend on the number of workers. Thus, the total execution time of one iteration in a \mbox{BSF-computer} with one master and $K$ workers can be calculated as follows:
\begin{equation}\label{Eq-T_K_a}
T_K = K(L + t_s) + t_w/K+K\cdot L+t_r+t_p,
\end{equation}
which is equivalent to
\begin{equation}\label{Eq-T_K_b}
T_K = 2L\cdot K+t_s\cdot K + t_r + t_p+ t_w/K.
\end{equation}
By reducing the right-hand side of the equation to the common denominator, we obtain
\begin{equation}\label{Eq-T_K}
{T_K} = \frac{{{K^2}(2L + {t_s}) + K({t_r} + {t_p}) + {t_w}}}{K}.
\end{equation}
Using equations \eqref{Eq-speedup}, \eqref{Eq-T_1} and \eqref{Eq-T_K}, we obtain the following equation for the speedup of BSF-program:
\begin{equation}\label{Eq-BSF-speedup}
a(K) = \frac{{K(2L + {t_s} + {t_r} + {t_p} + {t_w})}}{{{K^2}(2L + {t_s}) + K({t_r} + {t_p}) + {t_w}}}.
\end{equation}
Let us analyze $a(K)$ as a function depending on $K\geq1$. The function $a(K)$ takes the value 1 at  $K = 1$ which is concordant with the definition of the speedup and equation \eqref{Eq-speedup}. The function $a(K)$ is a continuous and positive definite function on the interval $[1; + \infty )$. Let us find the derivative of the function $a(K)$:
\begin{equation}\label{Eq-derivative}
a'(K) = \frac{{(2L + {t_s} + {t_r} + {t_p} + {t_w})({{{{{t_w}} \mathord{\left/
 {\vphantom {{{t_w}} K}} \right.
 \kern-\nulldelimiterspace} K}}^2} - 2L - {t_s})}}{{{{\left( {K(2L + {t_s}) + {t_r} + {t_p} + {{{t_w}} \mathord{\left/
 {\vphantom {{{t_w}} K}} \right.
 \kern-\nulldelimiterspace} K}} \right)}^2}}}.
\end{equation}
It follows from \eqref{Eq-derivative} that the derivative takes the value 0 at the point \mbox{$K=\sqrt {{t_w}/(2L + {t_s})}$}.
Moreover, the derivative takes positive values for $K<\sqrt {{t_w}/(2L + {t_s})}$ and negative values for $K>\sqrt {{t_w}/(2L + {t_s})}$. This indicates that the point $K=\sqrt {{t_w}/(2L + {t_s})} $ is the point at which the BSF-program speedup takes the maximum value. Thus, we may make a conclusion that the value $\sqrt {{t_w}/(2L + {t_s})}$ is the \emph{upper bound of the BSF-program scalability}:
\begin{equation}\label{Eq-scalability-bound}
K \leq \sqrt {\frac{{{t_w}}}{{2L + {t_s}}}}.
\end{equation}
Note that the upper bound of BSF-program scalability does not depend on the amount of time that the master is engaged in receiving and evaluating worker results.

One more important characteristic of a parallel program is parallel efficiency, calculated by equation \eqref{Eq-efficiency}. Let us estimate the efficiency of a BSF-program. Using equations \eqref{Eq-efficiency} and \eqref{Eq-BSF-speedup} we obtain
\[\begin{gathered}
  e = \frac{{2L + {t_s} + {t_r} + {t_p} + {t_w}}}{{{K^2}(2L + {t_s}) + K({t_r} + {t_p}) + {t_w}}} = \frac{{2L + {t_s}}}{{{K^2}(2L + {t_s}) + K({t_r} + {t_p}) + {t_w}}} +  \hfill \\
   + \frac{{{t_r} + {t_p}}}{{{K^2}(2L + {t_s}) + K({t_r} + {t_p}) + {t_w}}} + \frac{{{t_w}}}{{{K^2}(2L + {t_s}) + K({t_r} + {t_p}) + {t_w}}}. \hfill \\
\end{gathered} \]
Assuming $K \gg 1$, we have
\[\frac{{2L + {t_s}}}{{{K^2}(2L + {t_s}) + K({t_r} + {t_p}) + {t_w}}} \approx 0\]
and
\[\frac{{{t_r} + {t_p}}}{{{K^2}(2L + {t_s}) + K({t_r} + {t_p}) + {t_w}}} \approx 0.\]
Hence,
\begin{equation} \label{Eq-BSF-Efficiency-a}
e(K) \approx \frac{{{t_w}}}{{{K^2}(2L + {t_s}) + K({t_r} + {t_p}) + {t_w}}}
\end{equation}
for $K \gg 1$. Dividing both parts of the equation \eqref{Eq-BSF-Efficiency-a} by $t_w$, we receive the following approximate equation to estimate the parallel efficiency of a BSF-program:
\begin{equation} \label{Eq-BSF-Efficiency}
e(K) \approx \frac{1}{{1 + \left( {{K^2}(2L + {t_s}) + K({t_r} + {t_p})} \right)/{t_w}}}.
\end{equation}

\section{Conclusion}\label{Conclusion}
In this article, the new BSF (Bulk Synchronous Farm) model of parallel computations was introduced. The BSF model is intended for evaluating iterative numerical algorithms designed for distributed memory multiprocessors. One distinctive feature of the BSF model is the ability to evaluate the scalability of an algorithm in the early phases of its development. The structure of a BSF-computer was described. A BSF-computer includes one master-node and several worker-nodes connected by a communication network. The structure of a BSF-program was described. A BSF-program uses the SPMD (Single-Program-Many-Data) model according to which all the worker-nodes execute the same program but process different data. The execution of a BSF-program is divided into iterations. In each iteration, the master sends the orders to the workers; the workers execute the orders and send the results to the master; the master processes the results and checks the exit condition; if the condition is not satisfied, then  the master sends new orders to the workers, beginning the next iteration, otherwise, the calculations are stopped. A cost metric was constructed for BSF-programs. This metric offers the following simple estimation for the upper bound of scalability: \[K \leq \sqrt {\frac{{{t_w}}}{{2L + {t_s}}}},\] where $K$ is the number of worker-nodes, $L$ is the latency, $t_w$ is the time a BSF-computer with one worker-node needs to execute the order, and $t_s$ is the time needed to send an order to one worker-node, excluding latency.

A BSF-implementation of the \emph{NSLP algorithm} \cite{Sokolinskaya2017} was performed to validate the theoretical studies presented in this article. The NSLP algorithm is used to solve large-scale non-stationary linear programming problems. A BSF-implementation of the NSLP algorithm is described in article \cite{Sokolinskaya2017a}. The source code of this implementation is freely available on Github, at \url{https://github.com/leonid-sokolinsky/BSF-NSLP}. The results of the computational experiments presented in \cite{Sokolinskaya2017a} show that the BSF model accurately predicts the upper bound of scalability for the NSLP algorithm implemented as a BSF-program.

Future work concerning the BSF model includes the following directions. First, develop a formalism to describe BSF-programs through higher-order functions. Next, design and implement a BSF skeleton for the rapid development of BSF-programs in C++ using the MPI-library. Finally, validate the BSF model with different well-known iterative numerical methods.

\begin{flushleft}

\end{flushleft}
\end{document}